\documentclass[12pt]{article}
\usepackage{amsmath,amssymb,bm,epsf,epsfig,graphicx}
\usepackage{color}

\input epsf.sty
\newcommand{\VEV}[1]{\left\langle #1\right\rangle}

\newcommand{\tr}{\text{tr}}

\begin{document}
{}~ \hfill\vbox{
\hbox{OU-HET-770}
 \hbox{RIKEN-MP-60}
 }\break
\vskip 2.1cm

\centerline{\large \bf A Landscape in Boundary String Field Theory:}
\vspace{5mm}
\centerline{\large \bf New Class of Solutions with Massive State Condensation}

\vspace*{8.0ex}

\centerline{\large \rm Koji Hashimoto$^{(a,b)}$\footnote{Email: {\tt koji at phys.sci.osaka-u.ac.jp}},
Masaki Murata$^{(c)}$\footnote{Email: {\tt m.murata1982 at gmail.com}}
}

\vspace*{8.0ex}

\centerline { \it $^{(a)}$Department of Physics, Osaka University,
Toyonaka, Osaka 560-0043, Japan}
\centerline { \it $^{(b)}$Mathematical Physics Lab., RIKEN Nishina Center,
Saitama 351-0198, Japan}
\centerline { \it $^{(c)}$Institute of Physics AS CR, Na Slovance 2, Prague 8, Czech Republic}
\vspace*{2.0ex}

\vspace*{6.0ex}

\centerline{\bf Abstract}
\bigskip
We solve the equation of motion of boundary string field theory
allowing generic boundary operators quadratic in $X$, and explore 
string theory non-perturbative vacua with massive state condensation.
Using numerical analysis, a large number of new solutions are found. 
Their energies turn out to 
distribute densely in the range
between the D-brane tension 
and the energy of the tachyon vacuum.
We discuss an interpretation of these solutions as perturbative closed string states.
From the cosmological point of view, the distribution of the energies  can be regarded as the so-called landscape of string theory, as we have a vast number of non-perturbative string theory
solutions including one with small vacuum energy.
 \vfill \eject

\baselineskip=16pt

\section{Introduction}

As a non-perturbative formulation of open bosonic string, boundary string field theory (BSFT) \cite{Witten:1992qy,Witten:1992cr} was proposed as well as cubic string field theory (CSFT) \cite{Witten:1985cc}. 
In general, solutions of string field theories are quite important as they would provide
non-perturbative vacua of string theory, to look at the true capability of string theory.

Recently, the multiple D-brane solutions, which have greater energies than the trivial vacuum, were proposed \cite{Murata:2011ex,Murata:2011ep} in CSFT. 
It would have a significance equivalent to the proof of the 
original Sen's conjecture \cite{Sen:2004nf,Sen:1998sm,Sen:1999mh,Sen:1999mg,Sen:1999xm}, since
the D-brane creation is thought of as a necessary ingredient for a complete 
non-perturbative formulation of string theory. To climb up the SFT potential hill instead of
rolling down the hill to get to the tachyon vacuum, it is indispensable to treat the 
string massive modes.

After the construction of the analytic solution for tachyon condensation \cite{Schnabl:2005gv}, various analytic solutions in CSFT have been found \cite{Okawa:2006vm,Kiermaier:2007ba,Schnabl:2007az}.
In recent times, analytic forms of lump solutions \cite{Ellwood:2009zf,Bonora:2010hi} and multiple D-brane solutions were proposed.
In BSFT, as well, an analytic solution for tachyon condensation and lump solutions have been found 
\cite{Gerasimov:2000zp,Kutasov:2000qp,Kutasov:2000aq}. 

To solve the equation of motion of CSFT, we encounter the infinite-dimensional equation, which is hard to solve.
In fact, there are some subtleties of proposed solutions \cite{Erler:2011tc,Murata:2011ep,Takahashi:2011wk,Masuda:2012kt,Hata:2011ke,Hata:2012cy}.
On the other hand, there is a consistent truncation scheme which reduces BSFT to a standard field theory with a finite number of fields.
The BSFT action was constructed also for boundary interactions quadratic in the worldsheet
field $X$, 
corresponding to a subset of massive modes of open string \cite{Li:1993za}.

The purpose of this paper is to solve the equation of motion of the BSFT action for the quadratic boundary operators. 
In contrast to CSFT, only the tachyon field plays a significant role in the BSFT exact 
solution for tachyon condensation and the lump solutions such that the analysis is rather simple. 
For this reason,  it is natural to expect that one may obtain a new class of solutions by 
involving some more boundary operators, 
aiming at new string vacua and a construction of
a multiple-D-brane solution.

We adopt the BSFT action for
quadratic boundary interactions with arbitrary number of derivatives on the worldsheet
given in \cite{Li:1993za}, and solve the equation of motion numerically to find homogeneous static solutions. 
The condensation of the massive fields is taken care of to their all orders.
So the solutions are non-perturbative ones at the classical level of SFT, in the same sense as for the non-perturbative tachyon vacuum solutions of the BSFT.
We discover a large number of new solutions of BSFT. Interestingly,  
those energies turn out to be smaller than the D-brane energy.
Our analysis strongly suggests the existence of an infinite number of solutions.

We also find that an approximately uniform distribution of the energies of the solutions, which
suggests a relation to
closed string excitations at the tachyon vacuum.
Furthermore,
from a cosmological point of view, the distribution of infinitely many solutions is 
reminiscent of the so-called string landscape. 
It is intriguing that a solution with any small energy may be possible in BSFT, to reconcile the
cosmological constant problem.
We also find solutions with a part of Lorentz symmetry broken, which would serve as
a realization of
the old idea of spontaneous Lorentz symmetry breaking (and CPT breaking)
through SFT \cite{Kostelecky:1988zi,Kostelecky:1991ak,Kostelecky:1995qk}.

This paper is organized as follows.
In the next section we review the derivation of the BSFT action and derive the potential for the tachyon field and 
massive fields associated with generic quadratic boundary interactions
on the worldsheet.
From the potential, we obtain the equations of motion and solve them numerically in sec.\ref{sec:SolutionEOM}. 
We show plots of numerical results as well, to show the energy distribution of the solutions.
In sec.\ref{sec:Interpretation}, we present a possible interpretation of the solutions
as relevance to closed string states, and study properties of the solutions.
Finally, sec.\ref{sec:Discussion} is devoted to discussions.

\section{Review: the BSFT action}
We give a short review of boundary string field theory (BSFT) based on \cite{Witten:1992qy,Witten:1992cr}.\footnote{ See \cite{Shatashvili:1993ps,Shatashvili:1993kk} for relevant formulas and derivations.
For the supersymmetric formulation, see for example \cite{Kutasov:2000aq,Marino:2001qc,Niarchos:2001si,Ghoshal:2001bq}.
Recent proposals of the formulation includes \cite{Teraguchi:2006tb}.}
In addition, we summarize the derivation of the BSFT action for quadratic boundary interactions following 
\cite{Li:1993za}.

\subsection{Generic formulation of BSFT}
The dynamical variables of BSFT are boundary coupling constants $\lambda_i$ associated with the boundary operators $O^i$ of ghost number $1$. 
The BSFT action is given by the solution to the equation
\begin{equation}
\frac{\partial S}{\partial \lambda_i} 
= -\frac{1}2 T_{25} \int_0^{2\pi}\!\! d\theta \int_0^{2\pi}\!\! d\theta'\,
\langle
O^i(\theta) \, \{ Q_{\rm B}, O(\theta')\}
\rangle_\lambda\,.
\label{eq:dS}
\end{equation}
Here $T_{25}$ is the tension of the D25-brane, $O=\sum_i\lambda_iO^i$ and $Q_B$ is the BRST charge.
$\langle\cdots\rangle_\lambda$ denotes the correlation function in the two dimensional field theory on a unit disk, described by a bulk world-sheet action $S_{\rm bulk}$ with boundary interaction terms:
\begin{equation}
S_{\rm bulk} + \sum_i \lambda_i \int_0^{2\pi}\!\!  d\theta \, V^i(\theta)
\,.
\label{eq:WorldSheet}
\end{equation}
Here $\theta$ is the angle parametrizing the boundary of disk.
$V^i$ is the vertex operator associated with the open string state $\lvert V^i \rangle$:
\begin{equation}
\lvert V^i \rangle = b_{-1} \lvert O^i \rangle \,,
\end{equation}
where $\lvert O^i \rangle$ is as well associated with the boundary operator $O^i$.
One can show that the action $S$ satisfies the Batalin-Vilkovisky (BV) master equation and  so has a gauge symmetry \cite{Witten:1992qy}.

It was given in \cite{Witten:1992cr} to write down the action directly for general $V$ constructed only from matter fields. 
For general choice of $V$,
\begin{equation}
\{ Q, O(\theta) \} = \sum_{n=1}^\infty c\partial^n c\, F_n(\theta) \,,
\end{equation}
where $F_n$ are some matter operators.
Hence the ghost correlation functions appearing in \eqref{eq:dS} are of the form $\langle c(\theta)c\,\partial^n c(\theta') \rangle$.
Due to the form of the ghost correlation function
\begin{equation}
\langle c(\theta)c(\theta')c(\theta'') \rangle
= 2 \left(
\sin(\theta-\theta') + \sin(\theta'-\theta'') + \sin(\theta''-\theta)
\right)\,,
\label{eq:ccc}
\end{equation}
\eqref{eq:dS} can be written in the form of
\begin{equation}
\frac{\partial S}{\partial \lambda_i} 
= -T_{25}\int_0^{2\pi}\!\! d\theta \int_0^{2\pi}\!\! d\theta'\,
\langle
V^i(\theta)
\left(
A(\theta')+\cos(\theta-\theta')B(\theta') + \sin(\theta-\theta') C(\theta')
\right)
\rangle_{\lambda}\,,
\label{eq:dSABC}
\end{equation}
where $A, B, C$ are linear combinations of $F_n$.
The operator $A$ has the expansion in term of a basis $\{V_i(\theta)\}$ of matter operators,
\begin{equation}
A(\theta) = \sum_i \alpha_i V^i(\theta)\,.
\label{eq:Aalpha}
\end{equation}
Then the action is given by
\begin{equation}
S = -T_{25}\left(\sum_i \alpha_i\frac{\partial}{\partial\lambda_i}+g \right)Z\,,
\label{eq:Sgeneral}
\end{equation}
where $Z$ is the partition function of the world-sheet theory \eqref{eq:WorldSheet} and $g$ is a constant.

\subsection{BSFT action with generic quadratic boundary interactions}
Next, following Li and Witten \cite{Li:1993za},
we derive the BSFT action for the most general quadratic boundary operators.
Note that the quadratic part gives a free CFT on the worldsheet, so the truncation of string
theory to the one with generic quadratic boundary interactions is a consistent truncation
\cite{Kutasov:2000qp}. 
Solutions of the BSFT with the generic quadratic boundary interactions amounts
to solutions of the full theory.\footnote{This was the important observation for the proof of the
Sen's conjecture \cite{Sen:2004nf} by BSFT \cite{Gerasimov:2000zp,Kutasov:2000qp,Kutasov:2000aq}.}

The generic quadratic operators are\footnote{It is not 
necessary to normal order $V_{\mu\nu}^k$ since it doesn't have singularity for generic $u^{\mu\nu}(\theta)$ regular at $\theta=0$. 
In fact, in \cite{Li:1993za}, $V_{\mu\nu}^k$ was not normal ordered.
The normal ordered form, however, is useful to see how the action reduces to the one associated with $V=\frac{a}{2\pi}+\frac{1}{4\pi\alpha'}\sum_{\mu=0}^{25}u^{\mu}\eta^{\mu\mu}:X_\mu^2(\theta):$, 
so we adopt it in this paper. }
\begin{equation}
O = c(\theta) V(\theta)\,,~~
V(\theta) =
\frac{a}{2\pi} 
+ \frac{1}{4\pi\alpha'}\, :X_\mu(\theta)
\int_0^{2\pi}\!\! d\theta'\,u^{\mu\nu}(\theta-\theta')X_\nu(\theta'):\,.
\label{eq:BoundaryOp}
\end{equation}
Here $::$ stands for the normal ordering as 
\begin{equation}
:X_\mu(\theta)X_\nu(\theta'): = X_\mu(\theta)X_\nu(\theta')+2\alpha'\eta_{\mu\nu}
\ln |1-e^{i(\theta-\theta')}| \,.
\end{equation}
We use the standard closed string action
\begin{equation}
S_{\rm bulk} 
= 
\int d^2\sigma \sqrt{h}\, \left(
\frac1{4\pi\alpha'} 
h^{ab}\eta^{\mu\nu} \partial_a X_\mu \partial_b X_\nu
+ \frac{1}{2\pi} b_{ab}\nabla^a c^b
\right)
\,.
\label{eq:bulk}
\end{equation}
Here, $\{a,u^{\mu\nu}(\theta-\theta')\}$ is a set of boundary couplings and $h_{ab}$ and $\eta_{\mu\nu}=\text{diag}(-,+,+,\ldots)$ are the metrics on the world-sheet and the target space respectively.
Without loss of generality, we can assume $u^{\mu\nu}(\theta)=u^{\nu\mu}(-\theta)$.
Following the above terminology, $\{\lambda_i\}=\{a,u_k^{\mu\nu}\}$ with $u_k^{\mu\nu}=u_{-k}^{\nu\mu}=\int_0^{2\pi}d\theta\, u^{\mu\nu}(\theta)e^{-ik\theta}$, and 
\footnote{
Until just before \eqref{eq:SLiWitten}, we treat $u_k^{\mu\nu}$ for all $k\in Z$ as independent valuables when we take derivatives.
}
\begin{equation}
V^a(\theta) = \partial_a V = \frac1{2\pi}\,,~~
V^k_{\mu\nu} = \partial_{u_k^{\mu\nu}} V 
= \frac1{8\pi^2\alpha'} :X_\mu(\theta) \int_0^{2\pi}\!\! d\theta' e^{ik(\theta-\theta')} X_\nu(\theta'):\,.
\end{equation}
It is notable that the $V^k_{\mu\nu}$ can be formally expressed as a linear combination of quadratic local boundary operators $X_\mu\partial^rX_\nu$, which are the vertex operators corresponding to a constant field strength and to a set of massive modes of open string for $r>0$.
Since the world-sheet action is quadratic in $X$, we can solve this theory.
The boundary condition is deformed by the boundary interaction as
\begin{equation}
0 = \partial_rX^\mu(\theta) 
+ \int_0^{2\pi}d\theta'\,u^{\mu\nu}(\theta-\theta')X_\nu(\theta')\,,
\end{equation}
where $r$ is the radial coordinate of the unit disk. 
The Green's function satisfying this boundary condition is
\begin{align}
G_{\mu\nu}(z,w) &= \langle X_\mu(z,\bar{z}) X_\nu(w,\bar{w}) \rangle_\lambda \nonumber\\
&= - \frac{\alpha'}{2} \eta_{\mu\nu} \left(\ln|z-w|^2+\ln|1-z\bar{w}|^2\right)
- \alpha' A_{0,\mu\nu} \nonumber\\
&- \alpha' \sum_{k=1}^\infty 
\left(
A_{k,\mu\nu} (z\bar{w})^k + A_{-k,\mu\nu} (\bar{z}w)^k
\right)\,,
\label{eq:Green}
\end{align}
where
\begin{equation}
A_{0,\mu\nu} = -(u_0^{-1})_{\mu\nu} \,,~~
A_{k,\mu\nu} = \frac1{|k|} \eta_{\mu\nu} - \left(\frac{1}{|k|\eta+u_k}\right)_{\mu\nu}
~~\text{for}~~k\neq0 \,.
\end{equation}
Here $(1/(|k|\eta+u_k))_{\mu\nu}$ is the inverse matrix of $(|k|\eta^{\mu\nu}+u_k^{\mu\nu})$:
\begin{equation*}
\left(\frac{1}{|k|\eta+u_k}\right)_{\mu\rho}(|k|\eta^{\rho\nu}+u_k^{\rho\nu})=\delta_\mu^\nu \,.
\end{equation*}
Notice that the correlation function in \eqref{eq:Green} is evaluated with the bulk action \eqref{eq:bulk} with the boundary terms given by \eqref{eq:BoundaryOp}.
The partition function of the world-sheet theory is determined from the differential equation
\begin{align}
\frac{\partial}{\partial u_k^{\mu\nu}} \ln Z
&= -\frac{1}{8\pi^2\alpha'} \int_0^{2\pi}\!\! d\theta \int_0^{2\pi}\!\! d\theta'
e^{ik(\theta-\theta')}\langle :X_\mu(\theta) X_\nu(\theta'): \rangle_\lambda 
= \frac12 \, A_{k,\nu\mu} \,.
\end{align}
In the last equation, we have used \eqref{eq:Green} and $A_{k,\mu\nu}=A_{-k,\nu\mu}$.
By integrating this differential equation, we obtain the partition function
\begin{equation}
Z = \mathcal{N} \det(u_0)^{-1/2} e^{-a} 
\prod_{k=1}^{\infty} 
e^{k^{-1}\tr(\eta\cdot u_k)}
\det(1+k^{-1}\eta \cdot u_k)^{-1}\,.
\label{eq:PartitionFunction}
\end{equation}
Here $(\eta \cdot u_k)_\mu^\nu=\eta_{\mu\rho}u_k^{\rho\nu}$ and $\mathcal{N}$ is the normalization constant determined by demanding that the partition function reduces to $V_{26}$, the volume of the target space, for $a=u_k=0$.
The factor $\det(u_0)^{-1/2}$ in the partition function originates from the integration over the zero modes
\begin{equation}
\int d^{26}x\, e^{-\frac{1}{2\alpha'}u_0^{\mu\nu}x_\mu x_\nu}
= (2\pi\alpha')^{13} \det(u_0)^{-1/2}\,.
\end{equation}
Thus in the $u_0\to0$ limit, the factor $\det(u_0)^{-1/2}$ should be replaced by $(2\pi\alpha')^{-13} V_{26}$.
This fact determines $\mathcal{N}=(2\pi\alpha')^{13}$.
It is interesting to see how \eqref{eq:PartitionFunction} reduces to 
the partition function with the boundary interaction 
\begin{equation}
V_2 = 
\frac{a}{2\pi}+\frac{1}{4\pi\alpha'}\sum_{\mu=0}^{25}
u^{\mu}\eta^{\mu\mu}:X_\mu^2(\theta): \,,
\label{eq:V2}
\end{equation}
which was given in \cite{Witten:1992cr}.
The vertex operator $V_2$ is given by \eqref{eq:BoundaryOp} by choosing $u^{\mu\nu}(\theta)=u^\mu\eta^{\mu\nu}\delta(\theta)$ or choosing $u_k^{\mu\nu}=u^\mu\eta^{\mu\nu}$ for all $k$.
The so-called Weierstrass' product formula:
\begin{equation}
\Gamma(x) 
= \frac{1}{x} e^{-\gamma x} 
\prod_{k=1}^{\infty}\left( e^{k^{-1}x}(1+k^{-1}x)^{-1} \right)\,,
\end{equation}
where $\gamma$ is Euler's constant, leads to
\begin{equation}
Z \rvert_{u^{\mu\nu}(\theta)=u^\mu\eta^{\mu\nu}\delta(\theta)} = \mathcal{N} e^{-a}\prod_{\mu=0}^{25}\left(
\sqrt{u^\mu}\, e^{\gamma u^\mu} \Gamma(u^\mu)
\right) \,.
\end{equation}
This is nothing but the partition function for \eqref{eq:V2} given in \cite{Witten:1992cr}.

The remaining task to derive the BSFT action is to find $\{\alpha_i\}=\{\alpha_a,\alpha_k^{\mu\nu}\}$ and $g$.
By applying $\{Q_{\rm B},X_\mu(\theta)\}=c\,\partial_\theta X_\mu(\theta)$, one gets
\begin{equation}
\{Q_{\rm B}, O(\theta) \}
= c \partial_\theta c V(\theta)
- \frac1{4\pi\alpha'} : c\, X_\mu(\theta) \int_0^{2\pi}\!\! d\theta'
u^{\mu\nu}(\theta-\theta')\,c\, \partial_{\theta'}X_\nu(\theta'): \,.
\end{equation}
Substituting this into \eqref{eq:dS} and using the ghost correlation function \eqref{eq:ccc}, we obtain
\begin{equation}
A(\theta) 
= - V(\theta) 
- \frac{1}{4\pi\alpha'} : X_\mu(\theta) \int_0^{2\pi}\!\! d\theta'
\sin(\theta-\theta')u^{\mu\nu}(\theta-\theta') \partial_{\theta'}X_\nu(\theta'): \,.
\end{equation}
This implies
\begin{equation}
\alpha_a = -a\,,~~
\alpha_k^{\mu\nu} = \frac12k\left(u^{\mu\nu}_{k+1}-u^{\mu\nu}_{k-1}\right) - u_k^{\mu\nu}\,.
\label{eq:alpha}
\end{equation}
$g$ can be determined as follows.
According to \eqref{eq:dSABC},
\begin{align}
\frac{\partial S}{\partial a}
=-
\frac{T_{25}}{2\pi} \int_0^{2\pi}\!\! d\theta \int_0^{2\pi}\!\! d\theta'
\VEV{A(\theta')}
= T_{25} \sum_i \alpha_i \frac{\partial}{\partial \lambda_i} Z\,,
\end{align}
where we have used \eqref{eq:Aalpha} and the fact that $\theta$-integrals of $\cos(\theta-\theta')$ and $\sin(\theta-\theta')$ vanish.
On the other hand, using $\partial_aZ=-Z$ and \eqref{eq:alpha},
the derivative of \eqref{eq:Sgeneral} with respect to $a$ is
\begin{equation}
\frac{\partial S}{\partial a}
= T_{25} \left(\sum_i \alpha_i\frac{\partial}{\partial\lambda_i} + (g-1) \right)Z \,.
\end{equation}
Consequently, we obtain $g=1$ and
\begin{equation}
S = -T_{25}
\left(
a+1
+ \sum_{k=0}^\infty \left(\frac12k\left(u^{\mu\nu}_{k+1}-u^{\mu\nu}_{k-1}\right) - u_k^{\mu\nu}\right) \partial_{u_k^{\mu\nu}}
\right)Z \,.
\label{eq:SLiWitten}
\end{equation}

Since we are interested in homogeneous static solutions, we remove $-(u_0^{\mu\nu}/2)\partial Z/\partial u_1^{\mu\nu}$, which leads to the kinetic term of the tachyon field $T$, from \eqref{eq:SLiWitten}.
Hence the potential term is
\begin{equation}
U = T_{25}\int d^{26}x\,\, e^{-T}
\left(
T+1 
+ \sum_{k=1}^\infty \beta_k^{\mu\nu} \partial_{u_k^{\mu\nu}}
\right)
\prod_{k=1}^{\infty} 
e^{k^{-1}\tr(\eta\cdot u_k)}
\det(1+k^{-1}\eta\cdot u_k)^{-1}\,,
\label{eq:Potential}
\end{equation}
where
\begin{equation}
\beta_{1}^{\mu\nu} = \frac12u^{\mu\nu}_{2} - u_1^{\mu\nu} \,,~~
\beta_{k\geq2}^{\mu\nu} = \alpha_{k}^{\mu\nu}
= \frac12k\left(u^{\mu\nu}_{k+1}-u^{\mu\nu}_{k-1}\right) - u_k^{\mu\nu}\,.
\end{equation}
Setting $T=a+u_0^{\mu\nu}x_\mu x_\nu/2\alpha'$, one can reproduce \eqref{eq:SLiWitten} except the kinetic term.
In particular, $-u_0^{\mu\nu}\partial Z/\partial u_0^{\mu\nu}$ in \eqref{eq:SLiWitten} is obtained from $T$ in the parenthesis in the potential \eqref{eq:Potential}.

We further restrict our attention in the the case where $u_k^{\mu\nu}=u_k^\mu \eta^{\mu\nu}$.
Since the non-diagonal parts of $u_k^{\mu\nu}$ always accompany the other non-diagonal elements of $u_k^{\mu\nu}$, this restriction is consistent in the sense that $\partial V/\partial u_k^{\mu\nu}=0$ for $\mu\neq\nu$.
The potential for the tachyon field $T$ and the diagonal elements of $u_k$ is
\begin{align}
U = T_{25}V_{26}\,\, e^{-T}&
\left(
T+1
- \sum_{\mu=0}^{25}
\sum_{k=1}^\infty \beta_k^{\mu}\left(\frac{1}{k+u_k^\mu}-\frac{1}{k}\right)
\right) 
\prod_{k=1}^{\infty} \prod_{\mu=0}^{25}
e^{k^{-1}u_k^\mu}(1+k^{-1}u_k^\mu)^{-1}\,,
\label{eq:VDiagonal}
\end{align}
where
\begin{equation}
\beta_{1}^{\mu} = \frac12u^{\mu}_{2} - u_1^{\mu} \,,~~
\beta_{k\geq2}^{\mu} 
= \frac12k\left(u^{\mu}_{k+1}-u^{\mu}_{k-1}\right) - u_k^{\mu}\,.
\end{equation}
Here we focused on the homogeneous static fields and performed the integration $\int d^{26}x$.

\section{The solutions of the equations of motion}
\label{sec:SolutionEOM}

In this section, we solve the equations of motion derived from the non-perturbative
potential \eqref{eq:VDiagonal} of the BSFT.
Since we have infinitely many degrees of freedom, we adopt an approximation and solve them numerically.
We find a large number of solutions, and those solutions have peculiar properties.
First, we present the equations of motion, and then show how to solve them
numerically with an estimate of the validity of the approximation. Then finally
we study the peculiar properties of the energy distribution of the solutions.

\subsection{The equation of motion}
To find solutions of the equations of motion with non-vanishing $u_k$, we first solve $\partial U/\partial T=0$.
This gives two solutions, the first one is
\begin{equation}
T = \sum_{\mu=0}^{25}
\sum_{k=1}^\infty \beta_k^{\mu}\left(\frac{1}{k+u_k^\mu}-\frac{1}{k}\right)\,,
\label{Tsol}
\end{equation}
and the second one is $T=\infty$ with $u_k^\mu$ arbitrary.\footnote{ 
It is worth to note that both solutions lead $S=T_{25}Z$. 
This equality was also found in BSFT with the boundary interaction \eqref{eq:V2} \cite{Witten:1992cr}.
Since the difference between $S$ and $T_{25}Z$ was given by $dZ/d\ln\mu$, this equality implies that the solutions correspond to the conformal fixed points of the world sheet theory.}
Obviously the second one is the tachyon vacuum solution \cite{Kutasov:2000qp} since the potential energy \eqref{eq:VDiagonal} 
vanishes, and here we see the consistency with the truncated
solution of \cite{Kutasov:2000qp} explicitly.\footnote{ 
It is interesting that the second solution $T=\infty$
lets all the other equations of motion for $u_k^\mu$ be trivially satisfied, for any value of $\{ u_k^\mu \}$.
This is a generalization of the fact that at the tachyon vacuum any constant field strength 
of the massless gauge field is a degenerate 
solution. Probably this is related to the fact that there is no open string excitation
at the tachyon vacuum.}

In the following we consider only the first solution \eqref{Tsol}.
Substituting this solution back into \eqref{eq:VDiagonal} gives
\begin{align}
U &=  T_{25}V_{26}\,\, e^{-\sum_{\mu=0}^{25}f(u^\mu)}\,,
\label{eq:Vf}
\end{align}
where
\begin{align}
f(u) &=
\sum_{k=1}^\infty 
\left(
\beta_k\left(\frac{1}{k+u_k}-\frac{1}{k}\right) - k^{-1}u_k +  \log(1+k^{-1}u_k)
\right) \nonumber\\ 
&= 
\frac12 u_1+
\sum_{k=1}^\infty 
\left(
\beta_k\left(\frac{1}{k+u_k}\right) +  \log(1+k^{-1}u_k)
\right) 
\,,
\label{eq:f}
\end{align}
and
\begin{equation}
\beta_{1} = \frac12u_{2} - u_1 \,,~~
\beta_{k\geq2} 
= \frac12k\left(u_{k+1}-u_{k-1}\right) - u_k\,.
\end{equation}
Here, $u$ in $f(u)$ is the shorthand notation of $\{u_k\}$.
In the second line of \eqref{eq:f}, we have used $\sum_{k=1}^\infty(u_{k+1}-u_{k-1})=-u_1$.
Clearly, solutions of the equations of motion derived from the potential $U$ are the stationary points of $\sum_{\mu=0}^{25}f(u^\mu)$.
Now the potential energy at a stationary point is in the form of
\begin{equation}
U_* = T_{25}V_{26}\,\, e^{-\sum_{\mu=0}^{25}f(u^{n_\mu}_*)}\,,
\label{Uf}
\end{equation}
where $\{u_*^n\}$ is a complete set of solutions of $\partial f/\partial u=0$ labeled by $n$.

Our next task is to solve $\partial f/\partial u=0$.
Since this is an infinite dimensional equation, it is difficult to solve it analytically.
However, we first note that we find a solution
\begin{equation}
u_k^\mu =0 \,
\end{equation}
{for all $k\geq 1$ and $\mu$. This implies $T=0$ by using \eqref{Tsol}, so, the solution
is nothing but the trivial vacuum of the original D25-brane. It is important that 
the trivial D25-brane solution and the tachyon vacuum solution are allowed in our
generalized scheme, as a check of the consistent truncation of the BSFT.

To find nontrivial solutions with massive state condensation,  
we solve this $\partial f/\partial u=0$ numerically by truncating the fields as
\begin{equation}
u_k = 0\, ~~\text{ for}~~ k > k_c\,.
\end{equation}
The fact that the variation of $f(u)$ with respect to $u_k$ consists of only $u_{k}$ and $u_{k\pm1}$ implies $\partial f/\partial u_{k>k_c+1}=0$.
Hence the nontrivial equations we have to solve are
\begin{equation}
\frac{\partial f}{\partial u_k}=0\,,~~\text{for}~~k\leq k_c+1\,.
\label{eq:dfdu}
\end{equation} 
In general, there is no solution since the number of the equation is $k_c+1$ and is bigger than the number of degrees of freedom $k_c$.
We first neglect the stationary condition with respect to $u_{k_c+1}$ and find the numerical solution of
\begin{equation}
{\rm EOM}(k_c):~~
\frac{\partial f}{\partial u_k}=0\,,~~\text{for}~~k \leq k_c\,.
\label{eq:dfduTruncate}
\end{equation}
Let $v^{k_c,s}$ be the solution of EOM$(k_c)$, where $s$ is the natural number labeling  the solutions.
We sort solutions in ascending order in their values of $f$,  {\i.e.}
$f(v^{k_c,s_1})<f(v^{k_c,s_2})$ for any choice of $s_1<s_2$ in the set $\{s\}$.
If ${\partial f}/{\partial u_{k_c+1}}$ is sufficiently small at $u=v^{k_c,s}$, we can regard $v^{k_c,s}$ as the approximate solution of the whole system of equations \eqref{eq:dfdu}.

\subsection{Numerical solutions}
Let us show numerical solutions. We shall also explain how we can
make the solution accurate, in spite of the introduction of the effective cut-off $k_c$.
We find a large number of solutions having different energies. 

Since the energy is written in terms of the function $f$ as in \eqref{Uf}, first we study the numerical 
solutions in terms of the values of $f$, which can help the analysis easier.

We solve EOM$(k_c)$ numerically for $k_c=2,3, \cdots,6$.
The values of $f(u)$ and 
$|\partial f/\partial u_{k_c+1}|$ for the real solutions are plotted in Fig.\ref{fig:f_dfduAtKc}.
\begin{figure}[tbp]
\begin{center}
\begin{tabular}{c}
\begin{minipage}{0.5\hsize}
\begin{center}
\includegraphics[width=80mm]{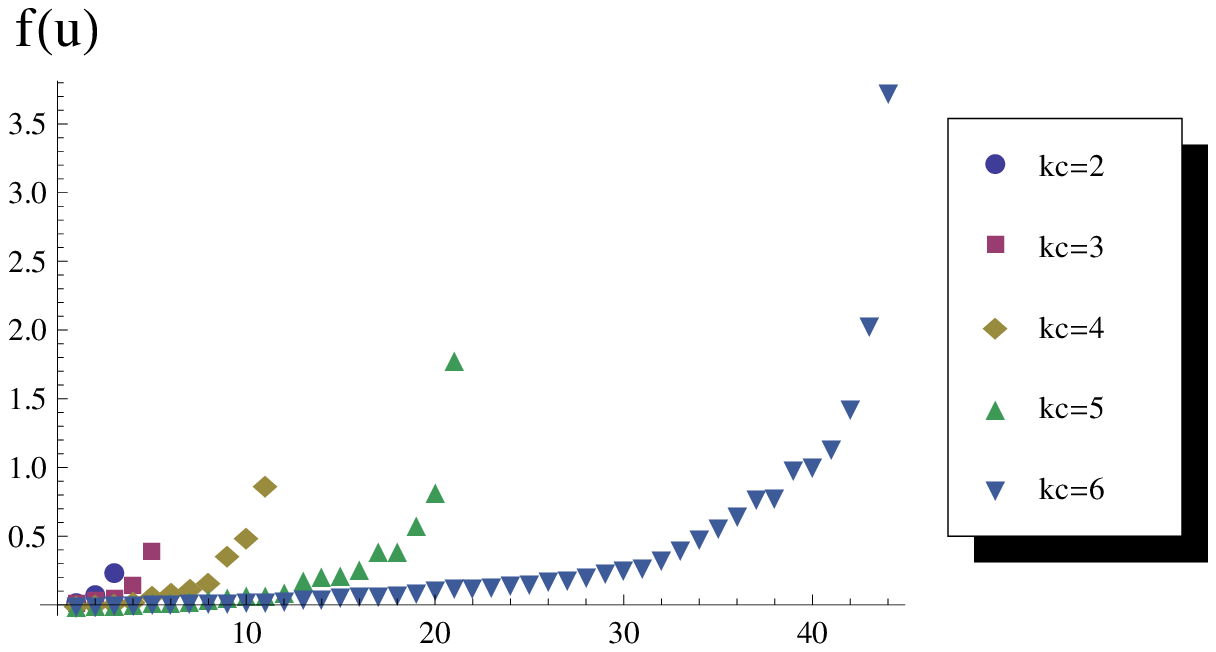}
\end{center}
\end{minipage}
\begin{minipage}{0.5\hsize}
\begin{center}
\includegraphics[width=80mm]{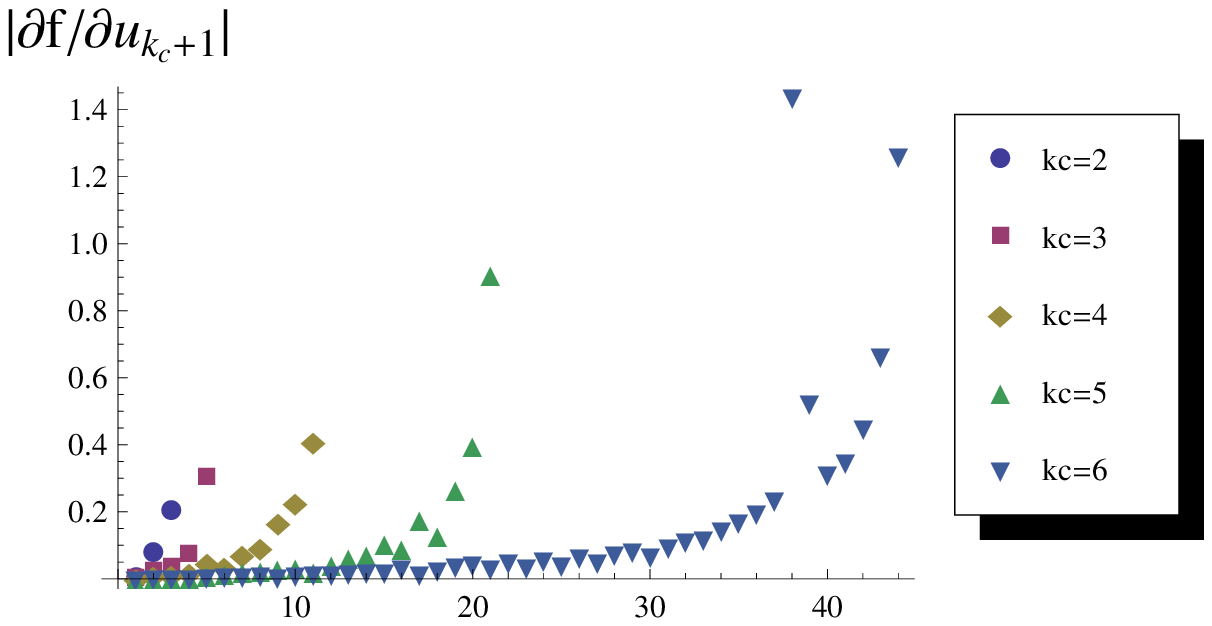}
\end{center}
\end{minipage}
\end{tabular}
\caption{
The left and the right graph show the values of $f(v^{kc,s})$ and $\lvert\frac{\partial f}{\partial u_{k_c+1}}(v^{kc,s})\rvert$ for $k_c=2,3,4,5,6$ respectively.
}
\label{fig:f_dfduAtKc}
\end{center}
\end{figure}
The number of real solutions are $3,5,11,21,44$ for $k_c=2,3,4,5,6$ respectively.
So the number of solutions diverges rapidly as we increase $k_c$.\footnote{
From this data, approximately, the number of the solutions $N(k_c)$ increases by a factor of 2 
when we increase $k_c$ by 1, so one can approximate $N(k_c)\propto 2^{k_c}$. However,
the data can be fit well also with $N(k_c)\propto k_c^3$. So we do not conclude whether the growth 
of the number of solutions is exponential or power-law-like. 
See the discussions at fig.~\ref{fig:IntervalMeanRatio}.}

At this stage, $\partial f/\partial u_{k_c+1}$ is not so small and is about the same magnitude as $f(u)$ itself. So, we cannot tell that these numerical solutions solve the full equation of motion. We shall improve the situation below.

For $k_c\geq7$, the computatios become much more complicated. 
Hence we take an alternative approach.
We begin with the $s$-th solution of EOM$(k_c)$, $v^{k_c,s}$. 
Using the Newton's method with the initial value
\begin{equation}
u_k = v_{k}^{k_c,s}\,~~\text{for}~~k \leq k_c\,,~~
u_{k_c+1} = 0\,,
\end{equation}
we obtain the solution of EOM$(k_c+1)$ which we call $w^{k_c,s,k_c+1}$.
In the same way, we can find the solution of EOM$(k_c+2)$ by applying the Newton's method  with the initial value
\begin{equation}
u_k = w_{k}^{k_c,s,k_c+1}\,~~\text{for}~~k \leq k_c+1\,,~~
u_{k_c+2} = 0\,.
\end{equation}
By iterating this procedure, we can obtain the solution of EOM$(L)$ for any $L\geq k_c$.\footnote{
It is, however, important to note that we get only a subset of all solutions of EOM$(L)$ since EOM$(L)$ is expected to have much more solutions than EOM$(k_c)$.}

The above procedure is shown to work well as an approximation, 
and provides a convergent solution solving the full equation of motion.
Let $w^{k_c,s,L}$ be the solution of EOM$(L)$ found by the iteration procedure which begins with $v^{k_c,s}$, the solution of EOM$(k_c)$. 
We found that as a function of $L$, $f(w^{k_c,s,L})$ converges to a nonzero finite value
as $L \to\infty$, while $\partial f/\partial u_{L+1}$ tends to vanish as ${\cal O}(1/L^2)$.
\footnote{
It is not clear from the first principle that vanishing of $\partial f/\partial u_{L+1}$ as ${\cal O}(1/L^2)$ is enough since the number of fields diverges as ${\cal O}(L)$.
The nontriviality of the equation of motion due to an infinite number of fields also appears in cubic string field theory.
We leave further study of this problem to future work.
}
In Fig.\ref{fig:F_DFDUkc6s5}, we show an example (among many solutions)
of the plots of $f$ and $\lvert\partial f/\partial u_{L+1}\rvert$ for $k_c=6,s=5$ as a function of $1/L$ and $1/L^2$  respectively.
In this example, $f\sim0.0071$ and $\lvert\partial f/\partial u_{L+1}\rvert\sim0.000017$ at $L=100$.
Therefore we claim that the equation of motion is approximately satisfied.
\begin{figure}[tbp]
\begin{center}
\begin{tabular}{c}
\begin{minipage}{0.5\hsize}
\begin{center}
\includegraphics[width=80mm]{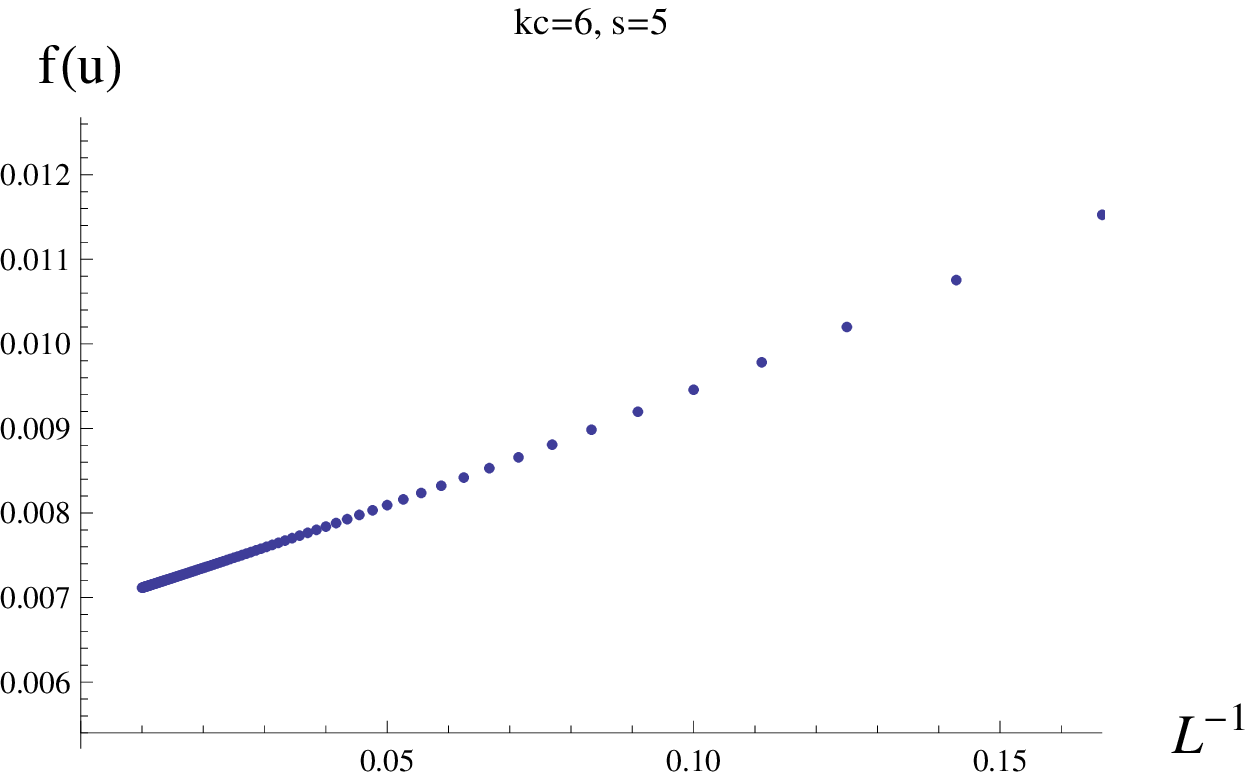}
\end{center}
\end{minipage}
\begin{minipage}{0.5\hsize}
\begin{center}
\includegraphics[width=80mm]{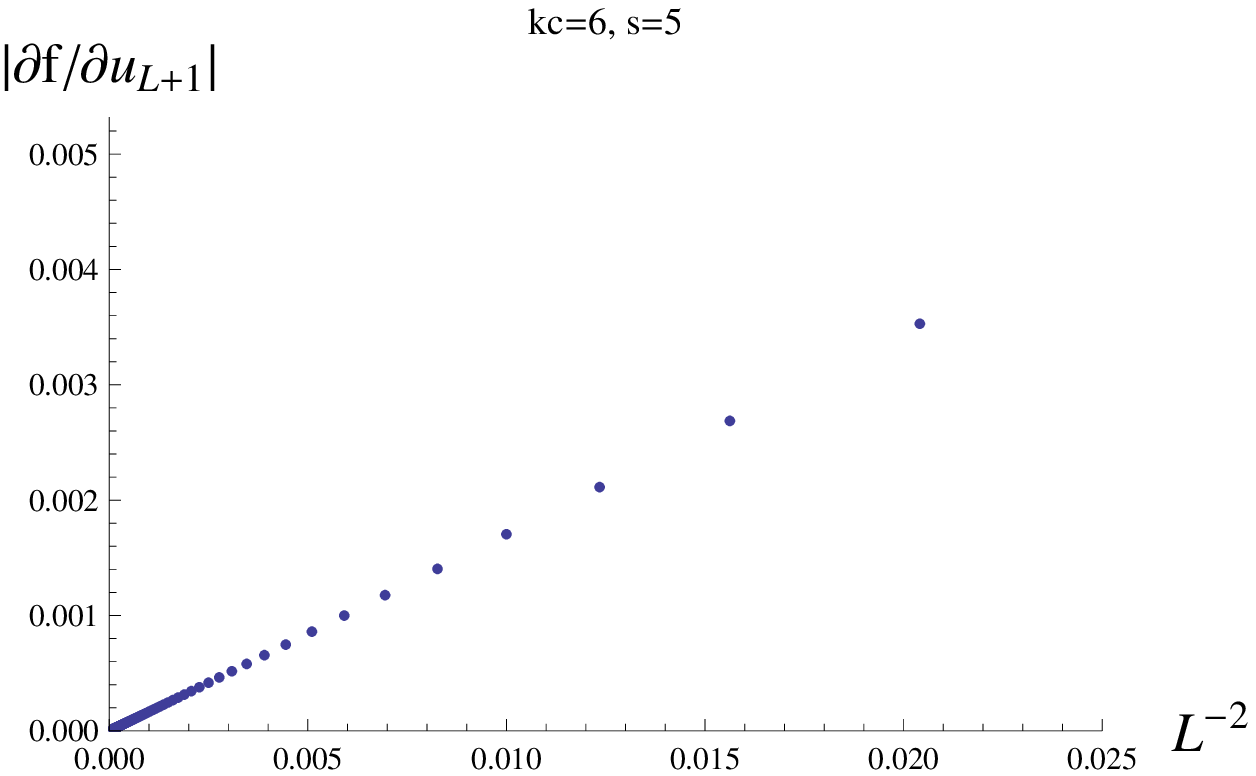}
\end{center}
\end{minipage}
\end{tabular}
\caption{
The plots of $f(w^{6,5,L})$ (the left graph) and $\lvert\frac{\partial f}{\partial u_{L+1}}(w^{6,5,L})\rvert$ (the right graph) as functions of $1/L$ and $1/L^2$ respectively.}
\label{fig:F_DFDUkc6s5}
\end{center}
\end{figure}
By fitting $f(w^{6,5,L})$ for large $L$, say $30\leq L\leq100$, to a quadratic function of $1/L$, it turns out to approach to a non-trivial value $0.006887$.

For any $k_c$ and $s$ within $2\leq k_c \leq6$, we extrapolate the values of $f$ at $L=\infty$, named $f_\infty(k_c,s)$, from the data for $30\leq L\leq 100$. 
Here the label $s$ of the solution is chosen in the same manner as before,
$f_\infty(k_c,s_1)<f_\infty(k_c,s_2)$ for any $s_1<s_2$.
The result is shown in Fig.\ref{fig:fAtLinfGrid}.
\begin{figure}[tbp]
\begin{center}
 \includegraphics[width=100mm]{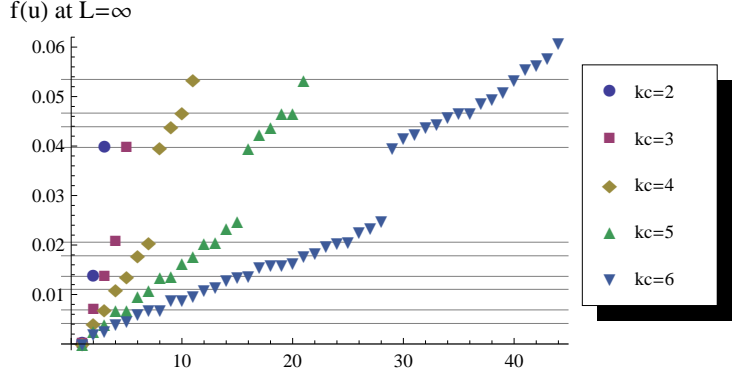}
 \caption{$f_\infty(k_c,s)$ extrapolated from the data for $30\leq L\leq 100$. Here we draw the horizontal lines in accordance with the values of $f_\infty(4,s)$.}
 \label{fig:fAtLinfGrid}
\end{center}
\end{figure}

The horizontal lines reveal that solutions of EOM(4) include ones of EOM(2) and EOM(3) and are included in ones of EOM(5) and EOM(6). 
For this reason, we expect that $\{w^{k_c,s,L=\infty}\}$ with fixed $k_c$ forms a subset of solutions of EOM$(k_c=\infty)$.

\subsection{Distribution of the energies of the solutions}
Based on the above results, we study the energies of the solutions.
We shall concentrate on Lorentz-invariant solutions for simplicity in this subsection.
Recall that the potential energy is written in terms of $f(u)$ as \eqref{eq:Vf}.
The Lorentz invariant solutions are given by choosing $u^\mu=u$ for all $\mu$.
Now the energy in units of $T_{25}V_{26}$ is given by $e^{-26f(u)}$ and so the extrapolated value at $L=\infty$ is $e^{-26f_\infty(k_c,s)}$.
The plots of the energies are given in Fig.\ref{fig:LorentzInvEnergy}.

We notice the following interesting features of the distribution of the energies of the solutions:
\begin{itemize}
\item[$\langle 1 \rangle$] {\it Almost uniform distribution of the energy spectrum.}

Although we have found a lot of solutions, the energy of those solutions do not overlap with each other, while tend to be uniformly distributed. This can be seen in Fig.\ref{fig:LorentzInvEnergy} 
as linear profiles of each dotted sector. 

\item[$\langle 2 \rangle$] {\it Energies below the D-brane tension.}

All the solutions we found have energies which take the values between the D-brane tension (normalized as 1.0 in Fig.\ref{fig:LorentzInvEnergy}) and the tachyon vacuum (no D-brane).

\item[$\langle 3 \rangle$] {\it Solution with a very small energy density.}

The lowest value of the energies among a fixed set decreases as $k_c$ increases. It suggests that
bringing up $k_c$ further would lower the lowest value of the possible range of the energy distribution.

\item[$\langle 4 \rangle$] {\it The presence of the ``desert".}

As seen from the plots in Fig.~\ref{fig:fAtLinfGrid}, there exists a ``desert"
in $0.25<f_\infty<0.4$ where there appears no solution up to $k_c=6$.
The desert, however, gets narrow as $k_c$ increases. 
For this reason, we expect that this desert is an artifact of finite $k_c$.
\end{itemize}

These features on energies are interesting, besides the surprising fact that we have obtained
a large number of solutions in string field theory. The large set of the 
solutions would serve as a ``string landscape."

It requires a possible interpretation in terms of string theory.
In the next section, we provide a possible interpretation of the solutions, with a detailed analysis of the intervals of the energies and the degeneracies of the solutions.

\begin{figure}[tbp]
\begin{center}
 \includegraphics[width=100mm]{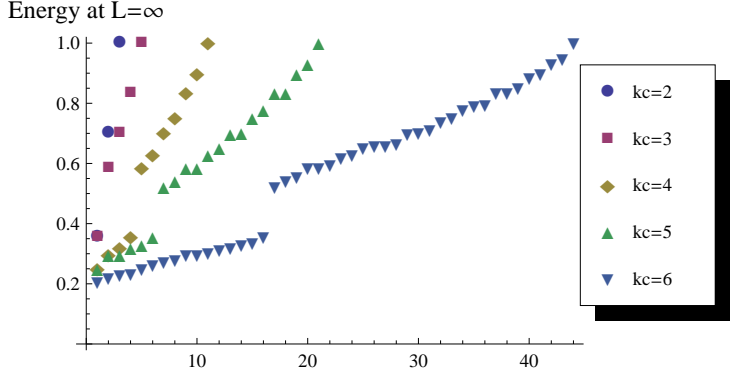}
 \caption{The plots of the energies for Lorentz invariant solutions in units of $T_{25}V_{26}$ extrapolated from the data for $30\leq L\leq100$. For every $k_c$, there is the solution whose energy is $1$, or $T_{25}V_{26}$. These are the trivial solutions $u=0$.}
 \label{fig:LorentzInvEnergy}
\end{center}
\end{figure}

\section{An interpretation of the solutions: closed strings}
\label{sec:Interpretation}

Closed strings, which should live at the tachyon vacuum as physical excitations, remain
a mystery in string field theory. There should exist closed string excitations, in particular
at the tachyon vacuum where the open string
degrees of freedom should go away along the disappearance of the original D-branes
{\it a la} Sen's conjecture. Now, 
a first look at the energy plot of Fig.~\ref{fig:LorentzInvEnergy} would suggest
a set of closed string states. The reasons are almost obvious: 
the plots appear to have a uniformly 
quantized energy levels (the feature $\langle$1$\rangle$ 
in the above list), and are below the D-brane tension, raising up from the 
tachyon vacuum (the feature $\langle 2 \rangle$).

In this section, we first present a detailed analysis on how the solutions are distributed,
and 
the existence of the degeneracy of the solutions. 
Then we discuss that both may serve as 
indirect evidence for our interpretation of our solutions as closed string states, 
but the true connection to the closed strings states
in the tachyon vacuum is yet to be unraveled.

\subsection{Properties of the solutions}
\subsubsection{Uniformity of the energy intervals}

We first analyze the distribution of the function $f$ for the solutions. 
To investigate the distribution of $f_\infty(k_c,s)$, 
we evaluate the intervals of pairs of $f_\infty(k_c,s)$:
\begin{equation}
g_\infty(k_c,s_1,s_2)=\frac1{s_2-s_1}\left(f_\infty(k_c,s_2)-f_\infty(k_c,s_1)\right) \,,
\end{equation}
for any pair $s_1 < s_2$.
Henceforth, we don't take into account the pairs separated by the desert.
The result is shown in Fig.\ref{fig:IntervalF}.
\begin{figure}[tbp]
\begin{center}
\begin{tabular}{c}
\begin{minipage}{0.5\hsize}
\begin{center}
\includegraphics[width=80mm]{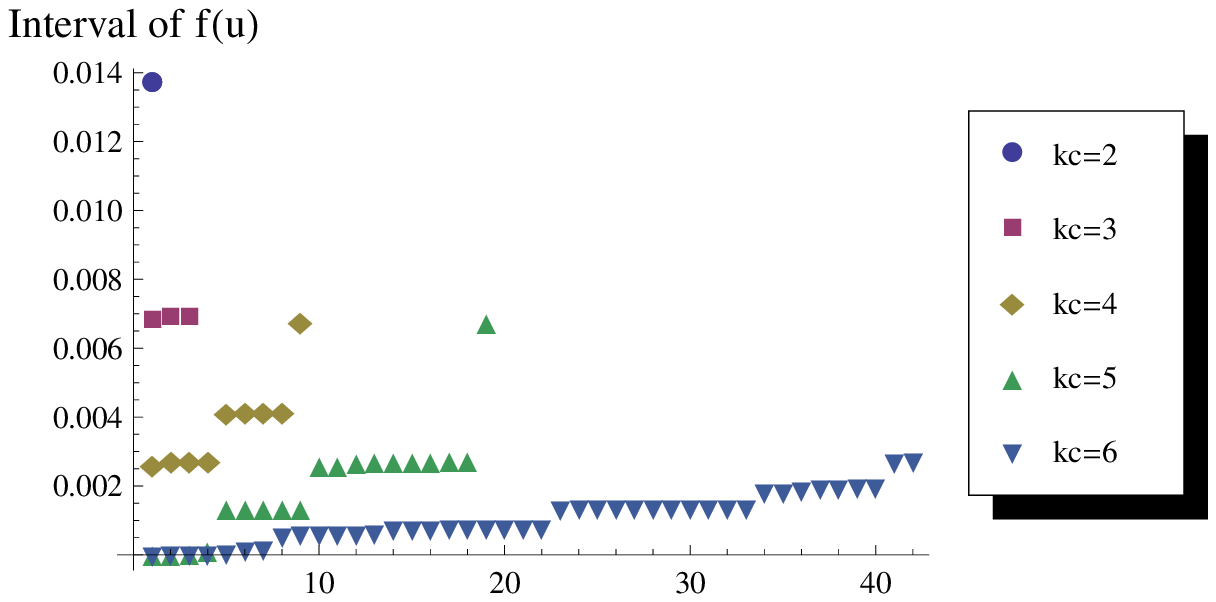}
\end{center}
\end{minipage}
\begin{minipage}{0.5\hsize}
\begin{center}
\includegraphics[width=80mm]{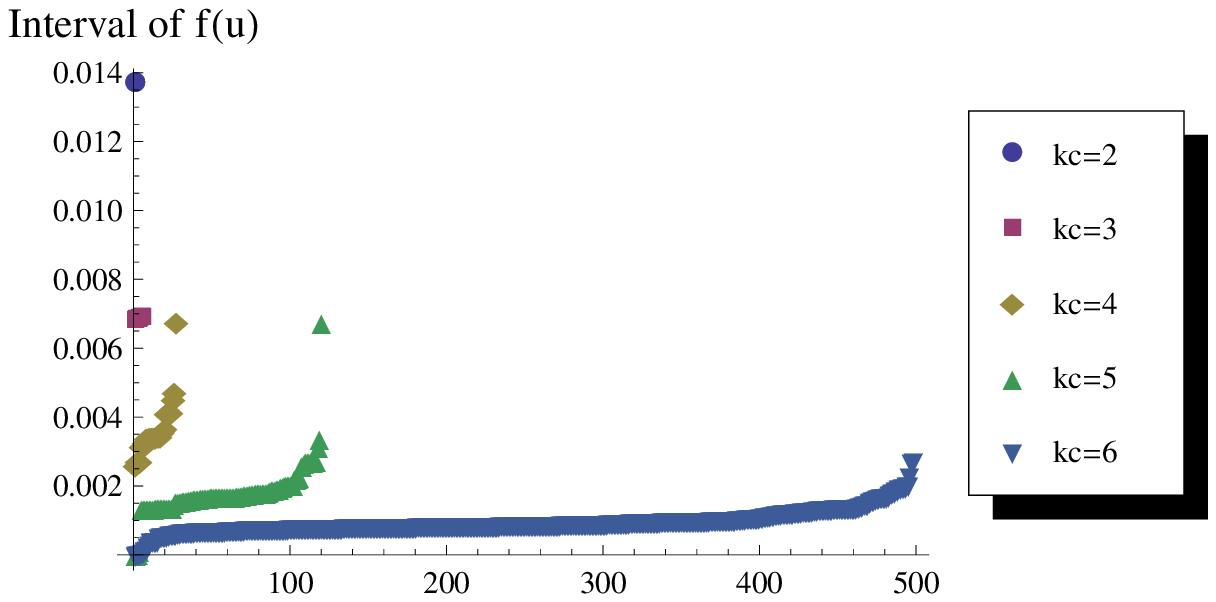}
\end{center}
\end{minipage}
\end{tabular}
\caption{The left graph shows the intervals of any paris of $f_\infty(k_c,s)$ next to each other, namely, $g_\infty(k_c,s,s+1)$, while $g_\infty(k_c,s_1,s_2)$ for general $s_1<s_2$ are shown in the right graph.}
\label{fig:IntervalF}
\end{center}
\end{figure}
It turns out that the intervals get smaller as $k_c$ increases.
This is the outcome of the increase of the number of the solutions.
It is important to note that the intervals become almost identical to each other.
This implies that the distribution of $f_\infty$ becomes uniform.

To see the intervals of the energies we define, in the same manner,
\begin{equation}
E(k_c,s_1,s_2) = \frac1{s_2-s_1} (e^{-26f_\infty(k_c,s_1)}-e^{-26f_\infty(k_c,s_2)})\,,
\end{equation}
where we don't take into account the pairs separated by the desert.
Again, the distribution of the energies is seemingly uniform except the desert.
See Fig.\ref{fig:IntervalEnergy}.
\begin{figure}[tbp]
\begin{center}
\begin{tabular}{c}
\begin{minipage}{0.5\hsize}
\begin{center}
\includegraphics[width=80mm]{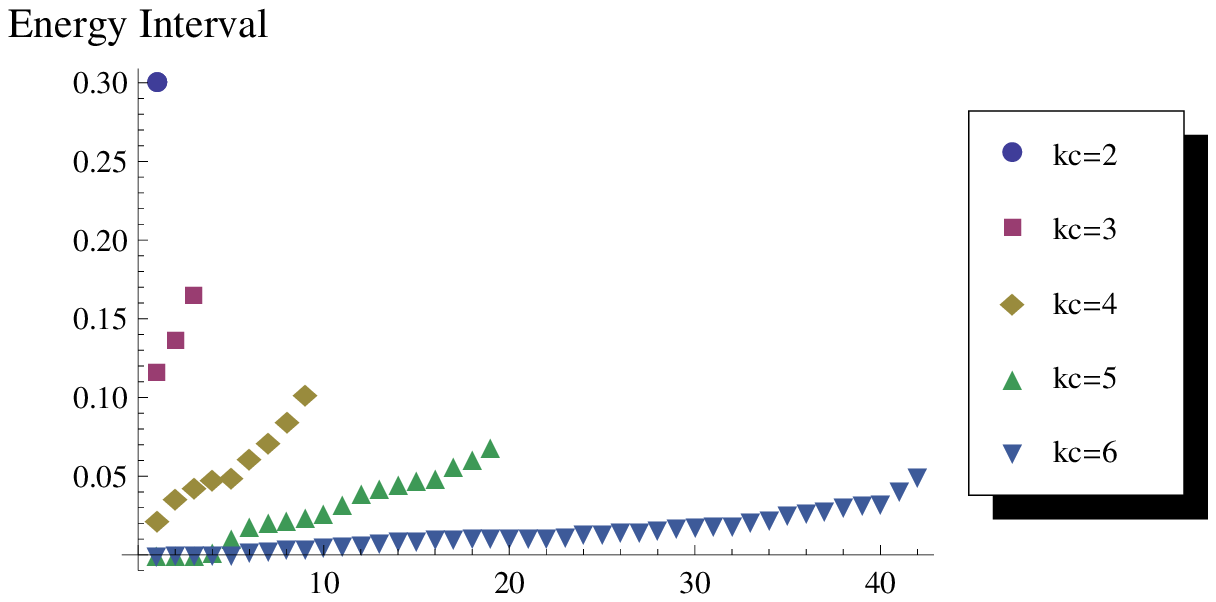}
\end{center}
\end{minipage}
\begin{minipage}{0.5\hsize}
\begin{center}
\includegraphics[width=80mm]{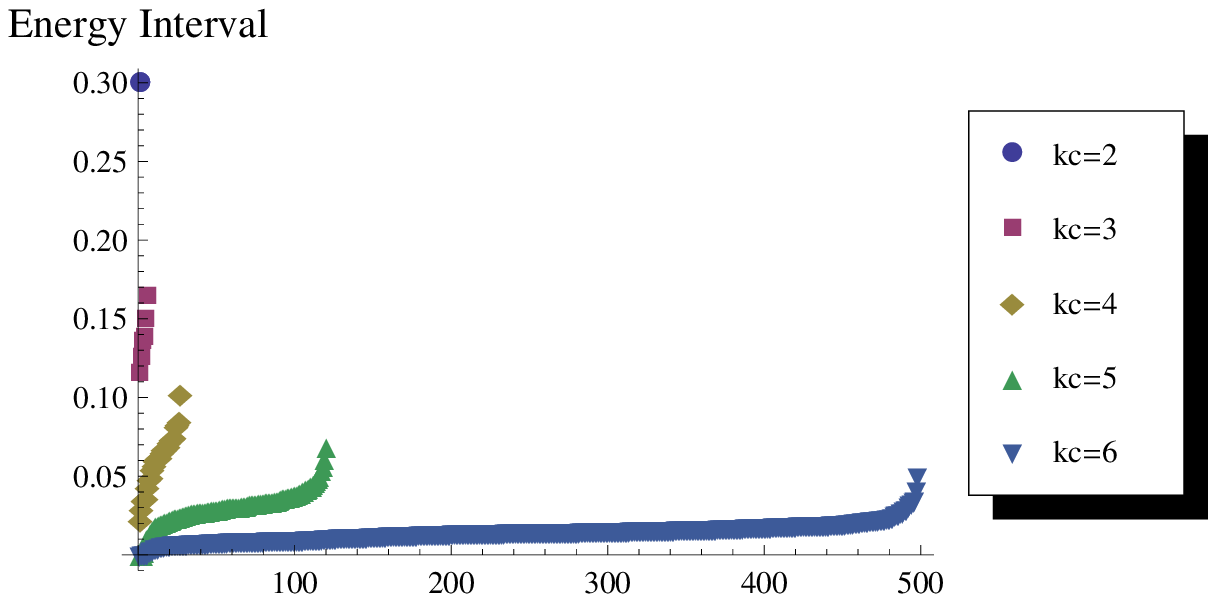}
\end{center}
\end{minipage}
\end{tabular}
\caption{The energy intervals in units of $T_{25}V_{26}$. The left graph shows $E(k_c,s,s+1)$, the energy intervals of any paris next to each other, while the intervals of any pairs are shown in the right graph.}
\label{fig:IntervalEnergy}
\end{center}
\end{figure}

So, as expected from a brief look at the energy distribution of Fig.~\ref{fig:LorentzInvEnergy},
it is indeed the case that the intervals between energy levels for different solutions are
approximately identical to each other, and the quantized distribution of the
energy levels are uniform.

The fact that the distribution of the energies is uniform
appears to be inconsistent with the uniformness of the distribution of $f_\infty$.
For the solutions we found, however, the values of $f_\infty$ are small so that $e^{-26f_\infty}\sim1-26f_\infty$ and so there is not much numerical difference between the intervals of $f_\infty$ and the ones of the energies.
At this stage, we can't precisely conclude whether $f$ or some function of $f$ like $e^{-26f}$ has a uniform distribution.
To obtain more precise prediction, we need to extend our computations to higher $k_c$ where we might get solutions with a sufficiently large value of $f$ to investigate which intervals are uniform.

The averaged interval is smaller for larger values of $k_c$. Fig.\ref{fig:IntervalMeanRatio} 
shows that
the averaged interval can be fit well as a function $2^{-k_c}$. 
This is consistent with the fact that the number of solutions grow as 
$2^{k_c}$ as stated earlier, assuming that finally in the $k_c\to\infty$ limit
the energy levels are uniformly distributed between the D-brane tension and the tachyon vacuum.

\begin{figure}[tbp]
\begin{center}
\begin{tabular}{c}
\begin{minipage}{0.5\hsize}
\begin{center}
\includegraphics[width=80mm]{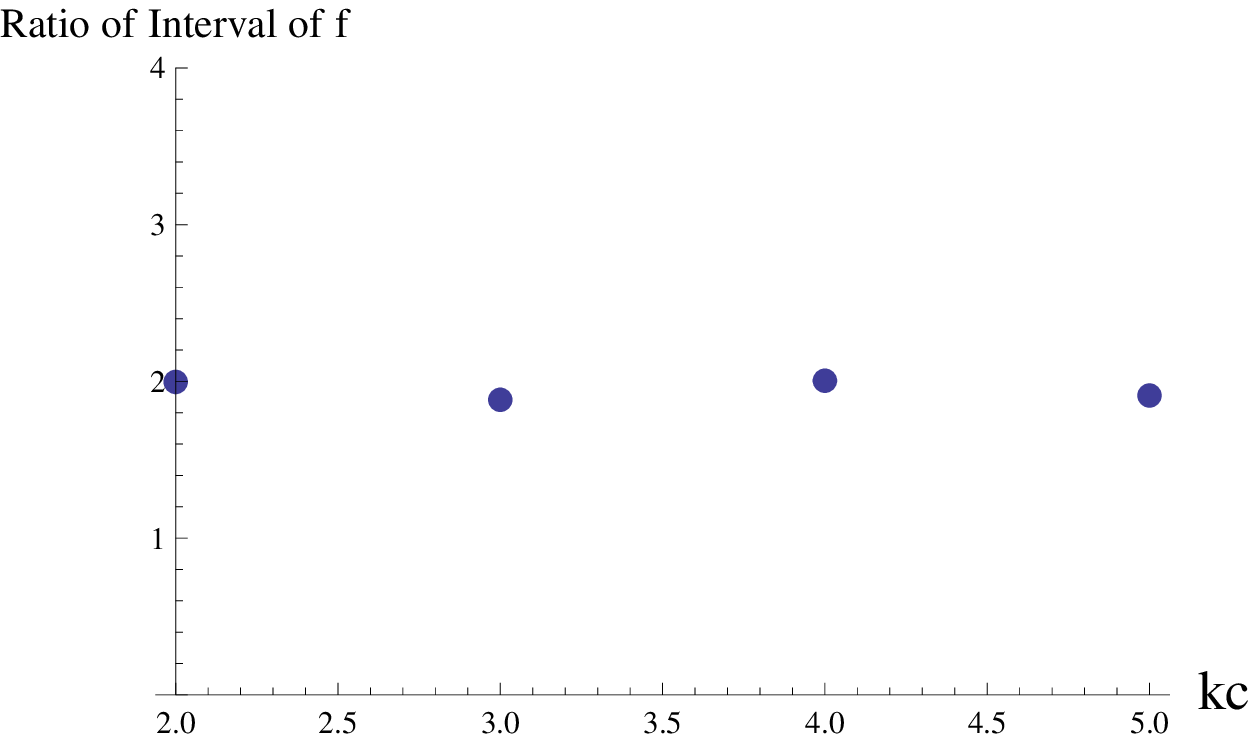}
\end{center}
\end{minipage}
\begin{minipage}{0.5\hsize}
\begin{center}
\includegraphics[width=80mm]{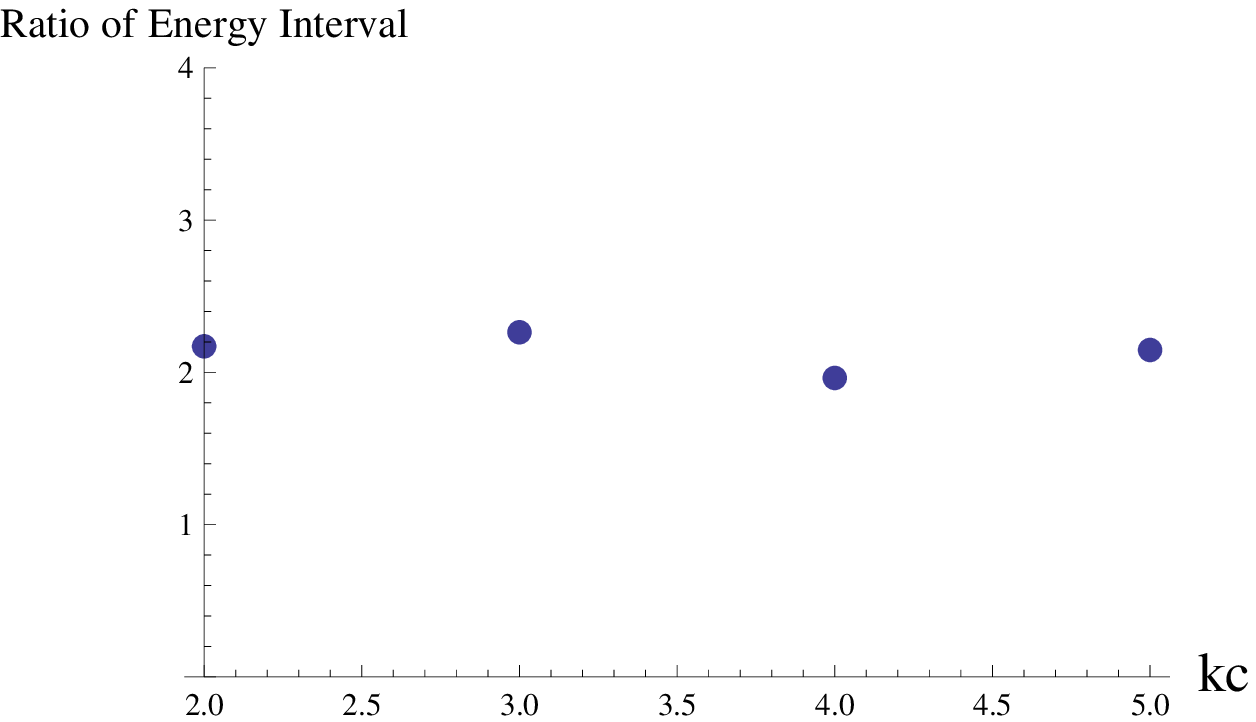}
\end{center}
\end{minipage}
\end{tabular}
\caption{
The ratio of the averaged intervals: the averaged intervals for $k_c$ divided by the one for $k_c+1$.
The left graph shows the ratio for $f$ and the right graphs shows the one for the energy.}
\label{fig:IntervalMeanRatio}
\end{center}
\end{figure}

\subsubsection{Degeneracy among the solutions}

Previously, we have considered only a solution with Lorentz invariance; we took all $u^\mu$ to
be equal to each other. 
In general, however, by choosing different solutions of $f(u)$ for different components of $u^\mu$, we obtain  solutions breaking Lorentz invariance.
Apparently, the energy spectrum degenerates since the equation of motion is still satisfied when we permute the components of $u^\mu$. 
This is an exact symmetry among solutions.

{Suppose we decompose 26 $u^\mu$ into two subsets. The $l$ components in the
first subset take $u^\mu=0$, while the other $26-l$ components take 
a single solution $u^\mu = u^\mu_*$. Obviously, we can find $26!/(26-l)! l!$
solutions with degenerate energy, from how we arrange the subsets. 
So, allowing a breaking of the Lorentz symmetry
provides a huge degeneracy in energies of the solutions.
Note that we can decompose 26 components further into a large number of subsets.

There is an additional form of an approximate degeneracy  
which comes with our numerical finding mentioned earlier, the approximate uniformity of
the distribution of the solutions in the $f_\infty$ space. 
Let us formally express solutions of EOM$(k_c=\infty)$ as $\{f_\infty(\infty,s)\}$.
The uniformity of the distribution implies
\begin{equation}
f_\infty(\infty,s) = s f_1\,, ~~\text{for}~~s=0,1,...\,,
\end{equation}
where $f_1=f_\infty(\infty,1)$.
It turns out that there are two types of solutions which have the same energy $e^{-2f_1}$:
\begin{align}
1:&~~ f(u^\mu)=f(u^\nu)=f_1\,,~~ \text{the other components}=0 \,, \nonumber \\
2:&~~f(u^\mu)=2 f_1\,,~~ \text{the other components}=0\,.
\label{eq:level2}
\end{align}
Therefore, the energy spectrum degenerates. This degeneracy adds up on the
exact degeneracy explained above. So, in total, we would have a huge degeneracy in energy,
among our solutions.

\subsection{An interpretation: closed string states?}

Let us discuss whether the properties of our BSFT solutions may allow an interpretation
as closed string excitations at the tachyon vacuum. As we will see, it is not conclusive.

First, let us see the uniform distribution of the energy of the solutions.
One should note that as the D-brane tension is inversely proportional to the string coupling constant
$g_s$, perturbative spectrum should appear infinitely dense if we normalize the D-brane 
tension to be the unity and take the perturbative string limit  $g_s\to 0$. So, the increase of the
number of the energy levels for larger $k_c$ is consistent with the interpretation that those are some perturbative
excitations of string theory.

The uniformity first looks as closed string energy levels, but note that 
our energy is the bulk energy of the BSFT solutions, while in closed string excited states
what is uniform is the closed string hamiltonian as a single-body problem.
Since we do not know why closed strings should be homogeneously distributed
in space, we lack a direct connection between closed strings and our BSFT solutions.

The standard 
closed string states have a huge degeneracy among states sharing the same energy.
As we saw above, we have found a similar degeneracy in our spectrum of the energies of the solutions.
\eqref{eq:level2}
is reminiscent of the closed string spectrum.
In fact, we can reproduce a part of the closed string degeneracies by our solutions,
and the representations under Lorentz transformation of a part of closed string spectrum by 
the following simple rule of a replacement: 
$f(u^\mu)=n f_1 \to A_{-n}^\mu=\alpha_{-n}^\mu\tilde{\alpha}_{-n}^\mu$.
For example, two solutions given in \eqref{eq:level2} correspond to $A_{-1}^\mu A_{-1}^\nu\lvert 0\rangle$ and $A_{-2}^\mu \lvert 0\rangle$, respectively.

There are, however, two important discrepancies: First, in our solutions, there seems to be 
no solution which amounts to the closed string state $A_{-1}^\mu A_{-1}^\mu \lvert 0 \rangle$.
Second, more notably, 
the degeneracy increases as the energy decreases, 
in contrast to the closed string spectrum, 
since $f(u)$ contributes to the energy in the form of $e^{-f(u)}$.
The latter problem would be complicatedly related to the infinitely large density
of energy levels in the $k_c \rightarrow \infty$ limit, which may require more exploration
of the solution space at larger $k_c$.

In sum, although the uniformity and the degeneracy are quite suggestive,
they are not sufficient to claim that our solutions are closed string states. It requires
further study for a conclusive interpretation.

\section{Discussion}
\label{sec:Discussion}

In the present paper, we have solved the equations of motion derived from the BSFT action  associated with general quadratic boundary operators. 
By means of numerical analysis, we have found a large number of solutions whose energies are uniformly distributed between the energy of the tachyon vacuum and the D-brane tension.
As the quadratic boundary interactions give a free worldsheet theory, the solutions we obtained
are solutions of the full string theory.

As we have discussed in sec.~\ref{sec:Interpretation},
our solutions are possibly related to non-trivial closed string excitations
at the tachyon vacuum (alternatively called the closed string vacuum).
In fact, it was discussed in \cite{Baumgartl:2004iy} that the non-local open string background implements shifts in the closed string background. 
By extending our analysis to larger $k_c$, 
we may make a progress on the interpretation of the solutions.

Our solutions in BSFT do not have any counterparts among known solutions in CSFT.
Although there are some suggestions on a possible relations between the two SFT's \cite{Coletti:2005zj},
it is difficult to see how our solutions may be mapped to CSFT. It would be interesting if
one can construct CSFT solutions sharing the properties with our solutions.
In addition, to gain insight on what our solutions mean, it is important to know
physical excitations around the found solutions. Whether the closed string excitations
actually can be identified would be the key point. Furthermore, an analogue of rolling
tachyon solutions in BSFT \cite{Sugimoto:2002fp, Minahan:2002if, Gibbons:2002tv}, 
and possible deformations of solutions in BSFT \cite{Hashimoto:2001rk}
would gain more insight. One of our Lorentz-violating solution resembles tachyon matter
solutions \cite{Sen:2002in}.

While we worked with the non-local boundary operators in this paper,
the BSFT action associated with the ``local'' boundary operators can be obtained by setting 
$u^{\mu\nu}(\theta)=\sum_{r=0}^st_r\delta^{(r)}(\theta)$.
Due to the contact divergence we need to introduce the short-distance cut-off $\epsilon$ and renormalize the coupling $a$ as $a'=a+\Delta a(t,\epsilon)$
\footnote{A different form of the renormalization using the so-called $\zeta$-function renormalization is discussed in \cite{Andreev:2000yn}.
} . 
With a proposed form of the counter term $\Delta a$ in \cite{Li:1993za}, the BSFT action $S_s(t_0,\ldots,t_s)$ turns out to  manifestly depend on $s$: $S_s(t_0,\ldots,t_s\to0)\neq S_{s-1}(t_0,\ldots,t_{s-1})$.
Since the counter term $\Delta a$ just shifts $a$, after we integrate out $a$, the action is independent of the choice of $\Delta a$. 
We found that this action again depends on $s$ in the above sense.
The physical interpretation of this fact is not clear and this is why we didn't work with the local boundary interactions.

In the introduction, we mentioned that massive mode condensation would be related to multiple-D-brane solutions in SFT. Although in sec.~\ref{sec:Interpretation} we discussed
our solutions may be interpreted as closed string states, they may still allow another
interpretation as multiple-D-branes.
We have not obtained a solution with energy larger than the original D25-brane,
it would not mean that our solutions are not multiple-D-brane solution. The reason is that 
since we are working in bosonic string theory, multiple-D-branes are unstable and would form
a bound state whose energy may be much smaller than just the multiple of the D-brane tension.
This kind of question can be answered only in superstring field theory, as superstring should have stable multiple-D-branes with energy protected by the BPS property, and further study is necessary.
It is, however, nontrivial to extend our analysis to superstring.
In fact to find the action is even more involved since the associated boundary operators are no longer quadratic in the matter operator $X^\mu$ contrary to the bosonic string
(see \cite{Hashimoto:2004qp} for an explicit treatment of massive states in super BSFT).

At a glance, our main result of fig.\ref{fig:LorentzInvEnergy} resembles a band structure of
electrons in materials. The fermion band structure is related to matrix models, and
indeed some matrix models represent tachyons and unstable D-branes (see
for example \cite{McGreevy:2003kb,Takayanagi:2003sm}). 
Suppose our solutions
with the massive field condensation are 
bound states of multiple unstable D-branes discussed above, then it is natural 
that matrix models appear as a low energy description of the multiple D-branes.
The excitations of the matrix models may look like a band structure. 
In general, the number of the excitations of the matrix model increases as the rank of the matrix $N$ increases. 
In this sense, $k_c$ plays a similar role as $N$. 
It is interesting to seek the counterpart of the ``desert" with given $k_c$ in the matrix models with finite $N$.

From the cosmological point of view, the distribution of infinitely many solutions is suggestive of the so-called landscape, which was found in superstring theory. Our solutions would be called as a landscape of BSFT.
\footnote{It is important to note again that to find the excitations around our solutions is important.
There would be tachyonic excitations.}
 Among its peculiar properties, 
it is intriguing that the lowest value of the energy decreases when we increase $k_c$.
If we take a large enough $k_c$, one may have a BSFT solution with very small cosmological 
constant. In sec.\ref{sec:Interpretation}, we also commented that we have 
solutions breaking Lorentz invariance.
This fact suggests that generic non-perturbative vacua may spontaneously break the Lorentz
symmetry. It reminds us of the original motivation of exploring non-perturbative vacua in
string field theories in late 80's: the spontaneous Lorentz and CPT violation 
\cite{Kostelecky:1988zi,Kostelecky:1991ak,Kostelecky:1995qk}. It would be interesting to
further explore our Lorentz-symmetry-violating
(metastable) vacua 
and its relevance to
the  important role in the baryon asymmetry of the universe.

\section*{Acknowledgements}
We would like to thank M. Schnabl and T. Erler for valuable discussions.
K.H. is supported in part by JSPS Grants-in-Aid  
for Scientic Research No. 23105716, 23654096, 22340069.
The research of M.M. was supported by grant GA\v{C}R P201/12/G028.

\end{document}